\documentclass[aps,prd,twocolumn,groupedaddress,nofootinbib,showkeys]%
{revtex4-1}

\usepackage{comment}

\usepackage[T1]{fontenc}
\usepackage{enumerate}
\usepackage{bm}
\usepackage{amsmath}
\usepackage{amssymb}
\usepackage{graphicx}

\newcommand{\C}{\Bbb C}
\newcommand{\T}{\Bbb T}
\newcommand{\MM}{\Bbb M}

\newcommand{\thorn}{\text{\TH}}
\newcommand{\omicron}{{o}}

\newcommand{\TT}{{\mathcal{T}}}

\newcommand{\scrif}{{\mathcal{I}^{+}}}

\newcommand{\scri}{{\mathcal{I}}}

\newcommand{\const}{\mathrm{const}}
\newcommand{\conj}{\mathrm{conjugate}}

\newcommand{\z}{{\mathfrak z}}

\newcommand\SB{{\sigma_{\rm B}}}

\newlength{\slashh}

\begin{document}

\title{Cut-continuity of twistor angular momentum}

\author{Adam D. Helfer}

\email[]{helfera@missouri.edu}
\affiliation{Department of Mathematics and Department of Physics \& Astronomy,
University of Missouri,
Columbia, MO 65211, U.S.A.}

\date{\today}

\begin{abstract}
Chen et al. argued recently that, in Bondi--Sachs space-times, the angular momentum at scri (null infinity) should vary continuously with the position of the cut (but not depend sensitively on its derivatives); they showed that this property was enjoyed by some definitions but not others.  I show here that the twistor definition has this continuity.  The argument is rather different from Chen et al.'s, with the invariant geometry of scri at the forefront.  
The flux, in the sense of the angular momentum emitted between two infinitesimally separated cuts, is calculated; this flux can be interpreted as the first variation of the angular momentum with respect to the cut.
Examining the second variation, one finds that the twistor definition, unlike most others,
responds to correlations in radiation between different asymptotic directions.
The twistor angular momentum is thus sensitive to qualitatively different structure in the asymptotic field than are the currently more used ones.
\end{abstract}



\maketitle

\section{Introduction}

The problems of elucidating both the physical content and the mathematical characterization of gravitational radiation occupied relativists for a good fraction of the last century.  
A burst of progress occurred around 1957--1962, with the ``sticky bead'' argument convincing hold-outs that the waves had physical significance, and then the papers of Bondi, Sachs and coworkers \cite{Bondi1960,BVM,Sachs1962a} considerably clarifying the mathematical structure involved.\footnote{It is not possible, in this space, to give credit to all those who made important contributions; see refs. \cite{Kennefick2007,HN2016} for many details.}  
This was based on insights which allowed the systematic and careful control of the gauge freedom appropriate to isolated systems.

One of their first and most powerful results was the identification of the energy-momentum of such a system, not just in its totality, but as a function of any Bondi retarded time parameter, what we now call a cut of Penrose's future null infinity $\scrif$ (scri plus).    
It was hoped one could get a parallel definition of angular momentum, but it became clear that that problem would require further conceptual advances.  Although Bondi, Metzner and Sachs (BMS) found what was naturally interpretable as a four-dimensional family of asymptotic translations at $\scrif$, there was no preferred extension to an asymptotic Poincar\'e group.  The usual foundations for the treatment of angular momentum were absent:  something new would have to be done.

There have since been many proposed approaches to this problem, and there is at present no consensus even on the general form of the solution (see refs. \cite{Szabados2009,CPWWWY,ADH2021b} for a review, some further recent references, and some comments).  Recently, Chen et al. \cite{CPWWWY} have argued that, whatever definition is adopted, it should not depend sensitively on the differential structure of the cut $\z$ at which it is evaluated.  More precisely, the angular momentum should vary continuously with uniform ($C^0$) changes in the location of the cut; one should not have also to control the derivatives of~$\z$.  This may be called {\em sup-norm continuity}.\footnote{So we are considering a map from the space of cuts to some sort of space of values of angular momentum, and we are asking if this is continuous when a relatively coarse topology is chosen for the {\em source } space.}

Although usually in physics the degree of differentiability of a quantity is not very significant, there are strong arguments for taking Chen et al.'s criterion seriously.  It would give angular momentum a kind of stability against perturbations which do not change the position of the cut much but make it crinkly, and it can be regarded as the ``next best thing'' to the existence of a flux; these are both highly desirable properties.  Chen et al. have shown that some proposed definitions (the Dray--Streubel BMS charges \cite{DS1984,Dray1985},
and the Chen--Wang--Wang--Yau (CWWY) spatial angular momentum \cite{CWWY})
are sup-norm continuous, but some others (most of those included in the Comp\`ere-Nichols family \cite{CN2021}) are not.

I will show here that the twistor definition of angular momentum \cite{ADH2007,ADH2021b} is sup-norm continuous.  This definition has a number of attractive features, particularly in providing physical interpretations of the quantities involved, being manifestly free of supertranslation ambiguities,
and having an intuitively satisfying notion of center of mass.
It perhaps would readily be expected to be sup-norm continuous, because it is built around Penrose's quasilocal twistor equations, which form an elliptic system and so have stability properties, but establishing the results is nonetheless technical.

Chen et al. bootstrapped their analysis by considering how definitions differed from the Dray--Streubel one (for which the result follows from the existence of a flux density\footnote{Unfortunately, there is no standardized terminology for the different flux concepts.  By a {\em flux,} I will mean the energy-momentum emitted in a strip between two infinitesimally separated cuts (possibly relatively supertranslated); a {\em flux density} will be a flux arising by integrating a three-form on $\scrif$ over the strip.}), reducing the problem to {\em ad hoc} questions about the differences.  However, the twistor approach is more delocalized than any of those and cannot be usefully regarded as a perturbation of them.
A direct analysis based on the twistor ideas is best, and this brings 
the invariant geometry of scri in the spin-coefficient formalism to the fore.

Along the way, explicit formulas for the twistor quantities at cuts supertranslated relative to a given Bondi system are derived; these will be more broadly useful.  It will follow from these that the twistorial definition extends sup-norm continuously to $C^0$ cuts.  The flux is easily computed.
It is natural to ask at this point whether the flux arises from a flux density.  It does not, and the argument gives us an important physical insight.

I note that, for a flux density of, in general, some quantity $A$ computed at cuts of $\scrif$ to exist, the second variation of the quantity with respect to the cut $\delta^2 A/\delta\z (\gamma_1)\delta\z(\gamma_2)$ (where $\gamma_1$, $\gamma_2$ are elements of the sphere of generators of $\scrif$) cannot have off-diagonal contributions (that is, non-zero values for $\gamma_1\not=\gamma_2$).  For the twistor angular momentum, we {\em do} find such off-diagonal contributions, and so no flux density can exist.  

But it is really the existence of the off-diagonal terms which is physically more important, for they signify that the twistor angular momentum responds to {\em correlations} between the gravitational radiation present in separate asymptotic directions.  Conversely, the existence of a flux density for (for example) the Dray--Streubel definition means it cannot involve such correlations.  There is a substantial qualitative difference in the structures of the radiation field detected by the twistor and Dray--Streubel definitions.

Finally, the work here clarifies the differing roles of the regularity properties of ``active'' and ``passive'' cuts.

To explain this more fully, some discussion of the 
significance of this 
regularity condition,
the difficulties facing proposals, and the twistor definition is in order.

\subsection{Angular momentum and regularity of cuts}

Chen et al. argue, in effect, that the angular momentum at a cut should not depend very strongly on any slight crinkliness of the cut --- the location of the cut should enter, but its derivatives should not.  I largely share this view.  Certainly, any definition which is claimed to be fundamental must either satisfy it or somehow provide a convincing explanation for not doing so.
It is worth noting, though, that this continuity depends on knowing the topology of the space in which the angular momentum takes values, and
most of the proposals require infinite-dimensional spaces whose topologies are usually not discussed.\footnote{The particular cases of angular momentum considered by Chen et al. are indexed by a choice of BMS vector field; in other words, their angular momentum takes values in the dual to the BMS algebra.  They analyze sup-norm continuity ``in the weak sense,'' that is, holding the BMS field constant.  This shows that the Dray--Streubel and CWWY definitions are sup-norm continuous in some reasonable topologies, but most of the Comp\`ere--Nichols ones cannot be.}

Sup-norm continuity is a highly desirable property, which
can be interpreted as being the ``next best thing'' to the existence of a flux for an angular momentum proposal.  (A proposal which does arise from a flux will automatically be sup-norm continuous.)  Moreover, sup-norm continuity of a quantity suggests it may extend continuously to cuts which are only $C^0$.  While this would not seem to be relevant for most physical modeling, it is of interest from the point of view of causal structure, where sets of low regularity naturally occur.  This is currently a particularly active area; see e.g. \cite{Steinbauer2022}.

But it
might at first seem surprising that {\em any} of the proposed definitions are {\em not} sup-norm continuous.
Most angular momentum proposals are framed in terms of integrals of spin-coefficients, spinor and tensor fields over the cuts --- how can such expressions be directly sensitive to the cuts' derivatives?  The contributions from invariantly defined tensor or spinor fields in the integrand, indeed, cause no difficulties.  But the derivatives of the cut itself can enter in two ways:  in defining the area element to be integrated, and 
in the computations of the shear (which actually depends on the second derivatives of the cut).  The area element contributes only one derivative, and that at first order, and if this were the only one it likely could be finessed by an integration by parts.  The more significant issues come from the shear.

The shear plays a key role in all definitions, and it would seem from the comments just made that in general we should expect angular momentum to be sensitive even to the $C^2$ structure.  However, the derivatives entering in the shear are in some sense gauge degrees of freedom, directly tied to the ``supertranslation'' problems.    From this point of view, sup-norm continuity of a definition is again desirable, as an indication that at least some gauge issues are being compensated.

\subsection{Origins and angular momentum}

In special relativity, we are used to representing angular momentum as a tensor field $M_{ab}(x)$ on Minkowski space.  If we try to construct a similar object in the asymptotic regimes of Bondi--Sachs space-times, we run into the difficulty that there is in general no preferred model Minkowski space to serve as a set of origins.  Indeed, the existence of supertranslation mismatches between different regimes at $\scrif$ can be viewed as a no-go theorem to this effect.

It is worthwhile spelling this out for a simple but important class of cases.  By a {\em regime} ${\mathcal R}$ at $\scrif$ I will mean a set between two nonintersecting cuts (or to the future of past of a single cut, or all of $\scrif$).  If the Bondi shear in a regime is purely electric and $u$-independent (where $u$ is the Bondi parameter), I will call the regime {\em Minkowskian}.   Then there is a Minkowski space $\MM({\mathcal R})$ whose points are the {\em good} (shear-free) cuts associated with the regime.\footnote{As a mathematical fiction, one extends the shear in the regime to all of $\scrif$, by keeping its $u$-independence,  to get the full Minkowski space of cuts.}
Equivalently, one could remove the shear in ${\mathcal R}$ by supertranslating, and for this reason such regimes are sometimes called {\em pure gauge} (as far as shear goes).

In a Minkowskian regime, there is a definition of angular momentum on $\MM({\mathcal R})$ which virtually all workers accept.  That is, there is a well-defined tensor field $M_{ab}^{\mathcal R}(x)$ on $\MM({\mathcal R})$, with origin-dependence of the correct form and the energy-momentum that of Bondi and Sachs.  (It is important to appreciate that this origin-dependence does not mean that the physical angular momentum varies in the sense of failing to be conserved.  It is just the fact that even a conserved angular momentum is conventionally represented by a tensor field.)  Definitions may provide information beyond this, but they should at least produce this much.

Problems occur when there are several Minkowskian regimes ${\mathcal R}_j$ whose good cuts are relatively supertranslated.  Then each has an angular momentum $M_{ab}^{{\mathcal R}_j}(x)$, which is a tensor field on $\MM({\mathcal R}_j)$, but
the supertranslation mismatches mean that there are no natural Poincar\'e motions identifying these different Minkowski spaces.  
The most direct statement of the supertranslation problem is to find a convincing way of relating or comparing the angular momenta $M_{ab}^{{\mathcal R}_j}(x)$ in spite of this.

Proposals to treat angular momentum at $\scrif$ must somehow deal with this.  The most common approaches vastly expand the set of origins, taking them to be all the (sufficiently smooth) cuts of $\scrif$. 
The angular momentum is then a function $M_{ab}(\z_{\rm act},\z_{\rm pas})$ of two cuts, an active one $\z_{\rm act}$ (at which we want to know the energy-momentum and angular momentum), and a passive one $\z_{\rm pas}$ (the choice of origin).\footnote{This terminology is due to Szabados.  For a conventional conserved angular momentum in special relativity, the dependence is purely passive.  (Any active dependence would imply non-conservation.)}  
In the notation of the previous paragraph, we want to compare angular momenta at different {\em active} cuts $\z_j$; since one cannot choose $\z_{\rm pas}$ to lie simultaneously in the different spaces $\MM({\mathcal R}_j)$, the idea is that one should allow, at least in principle, for $\z_{\rm pas}$ to be arbitrary.

One can indeed do this with proposals like the Dray--Streubel one.  
The issue is that none of the infinite-dimensional family of possible choices for $\z_{\rm pas}$ is strongly singled out.  And as (for each active cut) we vary $\z_{\rm pas}$, an infinite-dimensional family of cut-specific radiative data mixes in to $M_{ab}(\z_{\rm act},\z_{\rm pas})$.  
The proposal, built around the idea of using the BMS group as a formal parallel to the Poincar\'e group, succeeds so well in enforcing BMS homogeneity that it does not reduce the infinite-dimensional freedom.
Even for a fixed $\z_{\rm act}$, the angular momentum has an infinite-dimensional character.

Another concern is that, while the group-theoretic formal structure is parallel to that of special relativity, that formal parallel does not seem to lead to a physically satisfactory understanding of center of mass \cite{ADH2021}.  This point will be discussed in more detail in the next subsection.

All of these difficulties grow out of the idea of generalizing a Minkowski space of origins to the Bondi--Sachs context.  But there is another way of looking at special-relativistic angular momentum, a mathematically equivalent but quite different formalism, given by twistor theory.  And that formalism generalizes in a natural way to the Bondi--Sachs case.

\subsection{Twistors and angular momentum}

In Minkowski space, a {\em real twistor} can be thought of as a pair $Z=(\gamma,\pi_{A'})$ of a null geodesic $\gamma$ and a spinor $\pi_{A'}$ tangent to that geodesic.  This turns out to be intimately bound with angular momentum, for if we write that in spinor form
\begin{eqnarray}
  M_{AA'BB'}(x) =\mu_{AB}(x)\epsilon_{A'B'}+\mu_{A'B'}(x)\epsilon_{AB}\, ,
\end{eqnarray}
where $\mu_{AB}$ and its complex-conjugate $\mu_{A'B'}$ are the angular momentum spinors,
then the change-of-origin rule implies 
\begin{eqnarray}
  \mu^{A'B'}\pi_{A'}\pi_{B'}\text{ is constant along }\gamma\, .
\end{eqnarray}
This means that {\em the angular momentum can be regarded as a scalar-valued function}
\begin{eqnarray}
  A(Z) =2i\mu^{A'B'}\pi_{A'}\pi_{B'}
\end{eqnarray}
{\em on twistor space.}  (The factor $2i$ is conventional, as is the choice of the primed over the unprimed angular momentum spinor.)  The twistor $Z$ codes both some origin information ($\gamma$) and some choice of components ($\pi_{A'})$.  Allowing the twistors to vary, one can recover the field $\mu^{A'B'}$,  but there has been a basic shift in viewpoint:  {\em It is the null geodesics, rather than the events, which are taken to be the origins for angular momentum.}

Now let us turn to general relativity.  We may define a real twistor at $\scri$ as a null geodesic meeting $\scri$, together with a parallel-transported tangent spinor.  We thus have a space of (real) twistors, which is manifestly BMS-invariant.  (Notice, though, that it has a weaker structure than for Minkowski space, because of the supertranslational freedom to slide generators of $\scrif$ 
relative to one another.)  It also turns out that we can use Penrose's quasilocal kinematic construction to define an angular momentum twistor $A_\z(Z)$ at any cut.  The twistor $Z$ is arbitrary; it need not be specially related to the cut $\z$.

There is no difficulty at all in comparing the twistor angular momenta at different cuts, since they are functions on the same space:  one can form $A_{\z_1}(Z)-A_{\z_2}(Z)$.  (This is where the supertranslation problem would arise in conventional treatments.)
Also there are physically attractive twistor-derived definitions of spin and center of mass.  It turns out that twistors automatically compensate for any purely gauge effects (supertranslations).  All of this (we will see) depends only on the $C^0$ structure of the cut, and is sup-norm continuous.

How, though, are we to think of the twistor results?  How do we connect the function $A_\z(Z)$ with more conventional treatments?  This requires a little discussion:

(a) The purely twistorial formulation of the angular momentum itself is the function $A_\z(Z)$, which will be shown to be sup-norm continuous.  Mathematically, this is because the twistor construction naturally involves an angular potential $\lambda$ for the shear, this potential being sup-norm continuous.

(b) In special relativity, the center of mass is defined by the vanishing of the time-space components of the relativistic angular momentum (in a frame determined by the energy-momentum) --- a spatial vector, the mass-moment divided by the mass.  This is not a satisfactory view in general relativity.  For instance, in a Minkowskian regime ${\mathcal R}$, we do have such a description relative to the Minkowski space $\MM({\mathcal R})$, but if $\z_{\rm act}$ is a {\em bad} cut in ${\mathcal R}$, the center of mass, which will be a good cut, cannot simply be represented by a translation relative to $\z_{\rm act}$.  In a general circumstance we must expect the center-of-mass to be supertranslated relative to $\z_{\rm act}$, and therefore not represented simply by a three-vector.\footnote{This point seems not to be considered in most discussions, and in conventional approaches it is most common to see papers assume $\z_{\rm pas}=\z_{\rm act}$ {\em and also} apparently accept the three-vector of time-space components as the center of mass.  To try to deal with the issue, and keep the formal correspondence between the BMS and Poincar\'e groups,
it would be natural to consider
allowing $\z_{\rm pas}$ to vary to eliminate the time-space components.  But {\em that} would be expected to lead to severe ambiguities, since it amounts to constraining the infinite-dimensional freedom in $\z_{\rm pas}$ by the vanishing of three real numbers; this expectation has been verified
in the case of the Dray--Streubel definition \cite{ADH2021}.  If one knows one has good cuts, one can ``by hand'' restrict to these, but in the generic situation no resolution to this difficulty
has, to my knowledge, been offered.}

The twistor construction does exactly this.  The center of mass recovered from natural twistor formulas is in general supertranslated from $\z_{\rm act}$ by $\Re\lambda$, and automatically factors out any gauge issues.  In fact, what falls out of the twistor construction is that any
electric shear which may be present at $\z_{\rm act}$ is treated being due to a bad choice of cut.  In the case of a Minkowskian regime, even if $\z_{\rm act}$ is bad, the center of mass is given as a good cut.

The twistor center of mass is sup-norm continuous.

(c) Twistor theory treats the spin in formal parallel to the center of mass.  The result fits with an old observation of Newman and Winicour \cite{NW1974b}, which is that spin can be interpreted as a displacement of the center of mass into the complex.  Twistor theory assigns a three-vector ($j=1$) contribution to the spin, but it also interprets $M\Im\lambda$ as $j\geq 2$ components.\footnote{I will use $j$ (rather than $l$) for the multipole index, since some of the quantities will be spin-weighted.}

The twistor spin is also sup-norm continuous.

(d) To look for an interpretation of $A_\z(Z)$ parallel to the conventional field $\mu^{A'B'}$, note that in Minkowski space, for any point $p$, we would choose the twistors corresponding to null geodesics through $p$ to recover $\mu^{A'B'}(p)$; these geodesics would be those forming the good cut of $\scrif$ defined by $p$ --- they would meet that cut orthogonally.
In the Bondi--Sachs case, we no longer have compelling candidates for the Minkowski-space origins $p$, however.  What we may do is, as in the more conventional approaches, to fix a passive cut $\z_{\rm pas}$ and look at the twistors $Z$ whose geodesics meet $\z_{\rm pas}$ orthogonally.  That is, we consider $A_{\z_{\rm act}}(Z)$ for $Z$ meeting $\z_{\rm pas}$ orthogonally.
I will call this a {\em quasi\-conventional representation} of the twistor angular momentum; it is a generalization of $2i\mu^{A'B'}\pi_{A'}\pi_{B'}$, with the spinor $\pi_{A'}$ now corresponding to the different possible generators of $\scrif$ at which the twistors meet $\z_{\rm pas}$.

This quantity is a function on the sphere, which has not just $j=1$ terms but also $j\geq 2$ terms, essentially coding the same sorts of information as showed up in (b) and (c), above.  
Notice that in choosing the null geodesics {\em orthogonal} to $\z_{\rm pas}$ we have an explicit reference to the $C^1$ geometry of the {\em passive} cut, however; no additional properties of the {\em active} cut have been invoked.  The argument of Chen et al. applies only to motions of the {\em active} cut, and we will accordingly see that the quasi\-conventional representation of the angular momentum is sup-norm continuous.  Additionally, we will see the distinctions here between $\z_{\rm act}$ and $\z_{\rm pas}$ bring out a fuller understanding of their significance.

We can therefore think of the general-relativistic angular momentum as comprising two sorts of contributions:  a familiar, special-relativistic, $j=1$ term; and general-relativistic corrections $M\lambda$, which are $j\geq 2$.

What these results mean is that the twistor definitions of angular momentum, center of mass, spin and quasi\-conventional representation are robust and extend in a sup-norm continuous way even to $C^0$ active cuts.

\subsection{Outline and background}

The next section gives the basic definitions which will be used.  

Section III gives the details of the sup-norm continuity argument.  An important step is to work out the twistor fields at  cuts arbitrarily supertranslated with respect to the Bondi system, eqns. (\ref{twexn}), (\ref{twexp}); these results will be useful beyond this paper.  The next main step is to show the twistor fields are sup-norm continuous; it is at this stage that the hard-analysis technicalities come in.  The argument, though, just uses general results about elliptic operators; one does not need the sorts of ad-hoc inequalities of Chen et al.  

To finish the main continuity argument, we first identify the function space $A_\z$ lies in (it is naturally constructed from certain line bundles over the sphere of generators of $\scrif$).  Then its sup-norm continuity is established by  a series of parts-integrations to eliminate its apparent dependence on derivatives of $\z$.

With these preparations, most of the remaining results can be established almost by inspection.  Section IV shows the twistor of spin and center of mass are sup-norm continuous, and Section V that the quasi\-conventional representation is.  The first part of Section VI works out the flux; the second gives the results on flux density.  It is shown that because the twistorial angular momentum detects correlations in the gravitational radiation emitted in different asymptotic directions, it cannot admit a flux density.
The final section is given to discussion.

{\em Background and notation.}
This paper does not assume prior knowledge of twistor theory, but it also does not pretend to give a full account of the twistorial angular momentum; for that see \cite{ADH2007,ADH2021b}.

The paper does
assume a familiarity with two-component spinors and the spin-coefficient formalism at $\scrif$.  The divergence is taken to vanish, that is $\rho' =0$.

After some motivational discussion of twistors in Minkowski space, all computations are done at $\scrif$ in the spin-coefficient formalism, except that occasionally concepts are motivated by explicit references to the special-relativistic case.

The notation and conventions are as in Penrose and Rindler \cite{PR1984,PR1986}.  The usual smoothness assumptions at $\scrif$ (the manifold with boundary, and the conformal factor, being $C^4$, and the rescaled metric $C^3$) are adequate.
The speed of light is unity, and Newton's constant is $G$.

{\em I have indicated some important steps of the technical arguments in italics,} so the reader can easily find them.

\section{Definitions and set-up}

I first sketch the twistor treatment of angular momentum in special relativity, and then explain how this can be adapted to Bondi--Sachs space-times.

\subsection{Twistors and angular momentum in Minkowski space}

Special-relativistic twistors are the spinors of the conformal group of Minkowski space.  This group is $4-1$ covered by $SU(2,2)$, and twistor space $\T\simeq\C^4$ is the defining representation of the latter.  
Each twistor can be viewed as a spinor field, and also as a geometric structure, and this dual nature will be central.  

As a spinor field $\omega^A$, it is a solution to the {\em twistor equation}
\begin{eqnarray}\label{tweq}
  \nabla^{A'(A}\omega^{B)}=0\, .
\end{eqnarray}
Each solution has the form
\begin{eqnarray}\label{twsol}
  \omega^A(x) =\omega^A_0 -ix^{AA'}\pi_{A'}
\end{eqnarray}
for two constant spinors $\omega^A_0=\omega^A(0)$, $\pi_{A'}$.  These are called the coordinates of the twistor, and jointly denoted $Z^\alpha =(\omega^A_0,\pi_{A'})$.   

The norm on twistor space is
\begin{eqnarray}
  \Phi(Z) &=&\omega^A(x){\overline\pi}_A+\conj\\
    &=&\omega^A_0\overline{\pi}_A+\conj\label{norm}
\end{eqnarray}
(it is independent of position).  A twistor with $\Phi(Z)=0$ is called {\em real} or {\em null}.  Such a twistor vanishes on a unique null geodesic
\begin{eqnarray}
\gamma^{AA'}(s) &=& \frac{\omega_0^A{\overline{\omega_0}}^{A'}}{i{\overline{\omega_0}^{B'}}\pi_{B'}}
  +s{\overline\pi}^A\pi^{A'}
\end{eqnarray}
(which is real by virtue of the vanishing $\Phi(Z)$); conversely, a null geodesic $\gamma$ and a (parallel-transported) tangent spinor $\pi_{A'}$ determine a real twistor, and we may write $Z(\gamma ,\pi_{A'})$ for this.  Non-null twistors also have geometric interpretations, 
but we will not need these.  Also, twistors with $\pi_{A'}=0$ represent null geodesics at infinity, but we will not need to discuss these and will take $\pi_{A'}\not=0$.

In Bondi--Sachs space-times, the difficulties with treating angular momentum are associated with its position-dependence, and it is in this way that twistors offer an essentially new perspective.  

In special relativity, the angular momentum is a tensor field $M_{ab}$, changing as
\begin{eqnarray}
  M_{ab}(y+x) =M_{ab}(y) +P_ax_b-x_aP_b\, .
\end{eqnarray}
Write
\begin{eqnarray}
  M_{AA'BB'} =\mu_{AB}\epsilon_{A'B'}+\mu_{A'B'}\epsilon_{AB}
\end{eqnarray}
for the spinor decomposition, with complex-conjugate angular momentum spinors $\mu_{AB}=\mu_{(AB)}$, $\mu_{A'B'}=\mu_{(A'B')}$.  Then the component
\begin{eqnarray}
  \mu^{A'B'}\pi_{A'}\pi_{B'}
\end{eqnarray}
turns out to be {\em independent} of the position along a null geodesic $\gamma$ with tangent spinor $\pi_{A'}$.  {\em This means that the angular momentum may be regarded as a scalar-valued function on twistor space:}
\begin{eqnarray}
  A(Z) &=&A(\gamma,\pi_{A'}) =2i\mu^{A'B'}\pi_{A'}\pi_{B'}\, .
\end{eqnarray}
(The factor $2i$ is conventional.)  As the twistor varies, the different components of the angular momentum at the different points of space-time are determined, and indeed the energy-momentum is determined as well.

This is the twistor description of angular momentum.  It is, in special relativity, mathematically equivalent to the standard one, but it codes the information rather differently.  Each twistor $Z(\gamma ,\pi_{A'})$ contains both some location information (the geodesic $\gamma$) and some directional (or component) information (the spinor $\pi_{A'}$).  The angular momentum is a function, and simply a scalar-valued function $A(Z)$, on the space of twistors. 

Note especially that the origin-dependence of the angular momentum is coded very differently in the twistor formulation than in the conventional one.  This is what will resolve the difficulties in treating angular momentum in Bondi--Sachs space-times.

\subsection{Twistors for Bondi--Sachs space-times}

How can we carry over the twistorial treatment of angular momentum to $\scrif$ for Bondi--Sachs space-times?

There is an obvious candidate for the space of null twistors:  we take the pairs $Z=(\gamma, \pi_{A'})$, where $\gamma$ is a null geodesic meeting $\scrif$ and $\pi_{A'}$ is a parallel-transported tangent spinor.  This space is manifestly BMS-invariant.  If we could define the angular momentum twistor (and a few extra structures), we should have a definition.

Penrose's quasi-local twistor construction \cite{PR1986} suggests an approach.
It assigns to each cut $\z$ a four-dimensional complex vector space $\T (\z )$ of twistors, given as spinor fields $Z=\omega^A$ on $\z$, and also a candidate angular momentum twistor $A_\z(Z)$.  However, there are two difficulties with using just this information.  The first is that one does not have a clear way of determining which of these fields should count as real twistors, and knowing this is essential for extracting the angular momentum.  The second is that, in the presence of gravitational radiation, the spinor fields on the cuts do not ``integrate up'' to give well-defined fields on $\scrif$.  That means that there is no clear way of comparing the angular momenta at different cuts.  
One has an infinite-dimensional family of twistor spaces $\T (\z)$.

We can resolve these issues by exploiting the geometry of $\scrif$ to select certain preferred components of the twistor equation (\ref{tweq}) to enforce, and carrying over interpretive principles from Minkowski space.  What we will find is that the twistors exist invariantly, not as fields over all of $\scrif$, but as certain data on distiguished generators, which then give rise to quasilocal twistors when cuts are specified. 

\subsubsection{Null infinity and the spinor dyad}

We recall that Bondi coordinates at $\scrif$ comprise a {\em Bondi retarded time parameter} $u$ as well as suitable angular coordinates on the sphere of generators.  We will not need to actually pick coordinates on the sphere, and it will be shorter and conceptually better just to use a symbol $\gamma\in S^2$ to stand for such a generator.

Associated to a Bondi system is a null tetrad.\footnote{From now on, all quantities refer to the conformally rescaled metric, except when explicitly identified as special-relativistic.}    The vector $n^a$ is null and future-directed, points along the generators of $\scrif$, and is normalized to $n^a\nabla_a u=1$.  The vector $l^a$ is null, future-directed, and tangent to the $u=\const$ null hypersurfaces, normalized to $l^an_a=1$.  The vector $m^a$ is complex, null, tangent to the $u=\const$ cuts, antiholomorphic on $S^2$, orthogonal to $l^a$ and $n^a$, and normalized to $m^a{\overline m}_a=-1$.
There is a freedom to alter the phase of $m^a$, which is naturally handled by working with the calculus of spin-weighted quantities.  The spin-weighted derivatives in the $l^a$, $m^a$, ${\overline m}^a$, and $n^a$ directions are $\thorn$, $\eth$, $\eth'$, $\thorn'$, respectively.

This basis determines, uniquely up to an overall sign, a spinor dyad $\omicron^A$, $\iota^A$ with $l^{AA'
}=\omicron^A\omicron^{A'}$, $n^{AA'}=\iota^A\iota^{A'}$, $m^{AA'}=\omicron^A\iota^{A'}$ and $\omicron_A\iota^A=1$.

It is important to note that the direction of $\iota^A$ is determined invariantly (along the generators of $\scrif$), but $\omicron^A$ is sensitive to the choice of cuts.  If we supertranslate to a system with cuts $u=\z +\const$, where $\z$ is a function of angle only, then we will have
\begin{eqnarray}\label{transom}
  \omicron_\z^A &=&\omicron^A +(\eth\z)\iota^A\, ,
\end{eqnarray}
where the postscript $\z$ indicates a quantity (in this case, one of the basis spinors) adapted to the supertranslated system of cuts.

Finally, let me note the spinor conventions for components.  These are set by
\begin{eqnarray}
  \omega^A &=&\omega^0\omicron^A +\omega^1\iota^A\, .
\end{eqnarray}
I will shortly introduce a downstairs spinor field ${\tilde\pi}_{A'}$.  One has
\begin{eqnarray}
  {\tilde\pi}_{A'} &=&{\tilde\pi}_{1'}\omicron_{A'}-{\tilde\pi}_{0'}\iota_{A'}
\end{eqnarray}
owing to the lowering convention ${\tilde\pi}_{A'}={\tilde\pi}^{B'}\epsilon_{B'A'}$.  
Note that the transformation rule (\ref{transom}) implies
\begin{eqnarray}
  \omega^0_\z &=&\omega^0 \label{omzch}\\
    \omega^1_\z &=&\omega^1-(\eth\z)\omega^0 \label{omonech} \\
    {\tilde\pi}_{0'}^\z &=&{\tilde\pi}_{0'}+(\eth'\z){\tilde\pi}_{1'}  \label{pizeroch}\\
  {\tilde\pi}_{1'}^\z &=&{\tilde\pi}_{1'} \, ,
\end{eqnarray}  
where
\begin{eqnarray}
  \omega^A &=& \omega^0_\z \omicron^A_\z +\omega^1_\z\iota^A_\z\, ,\label{omz}\\
 {\tilde\pi}_{A'} &=& {\tilde\pi}_{1'}^\z \omicron_{A'}^\z -{\tilde\pi}_{0'}^\z\iota_{A'}^\z\label{piz}
\end{eqnarray}
are the decompositions relative to $u=\z+\const$ cuts.

\subsubsection{Twistor data}

We have seen that in Minkowski space, twistors can be viewed as spinor fields $\omega^A$ satisfying the twistor equation (\ref{tweq}), and each is determined by data $\omega_0^A=\omega^A(0)$ (the value of the field at the origin) and $\pi_{A'}$, where $\nabla_{AA'}\omega^B =-i\epsilon_A{}^B\pi_{A'}$ (from eqs. (\ref{tweq}), (\ref{twsol})).  
Certain geometrically distinguished components of these relations will be carried over to $\scrif$ to determine the twistors there.

Let me start though with some relations which do {\em not} reflect this distinction very strongly.  The first pair are the {\em quasilocal twistor equations}
\begin{eqnarray}
  \eth'\omega^0&=&0  \label{tweqa}\\
  \eth\omega^1 &=&\SB\omega^0\, ,\label{tweqb}
\end{eqnarray}
where $\SB$ is the Bondi shear.  (These are the two components of the twistor equation only involving derivatives tangential to the cut.)  The second pair define ${\tilde\pi}_{A'}$ (the conformally extended counterpart of the Minkowskian $\pi_{A'}$)\footnote{In some papers, especially those working exclusively at $\scrif$, this field is simply written $\pi_{A'}$.} via
\begin{eqnarray}
  {\tilde\pi}_{0'} &=& i\left(\eth'\omega^1-\rho \omega^0\right)\\
  {\tilde\pi}_{1'} &=& i\eth\omega^0\, .
\end{eqnarray}
(Here $\rho$ is one of the spin-coefficients, the convergence of $l^a$; recall that $\rho'$, the convergence of $n^a$, is taken to vanish.)
The quasilocal twistor equations (\ref{tweqa}), (\ref{tweqb}) form an elliptic system on any cut, with a four-complex-dimensional set of solutions.  These may be specified by the values $(\omega^0,\omega^1,{\tilde\pi}_{0'},{\tilde\pi}_{1'})$ at any point of the cut.  It is such data, carefully chosen, which will form our twistor space.

Now let us bring in the distinctions made by the geometric structure more strongly.  Note that the component $\omega^0 =-\omega^A\iota_A$ is BMS-invariantly defined (as a spin-weighted quantity).  We also see that the equation (\ref{tweqa}) is independent of $\omega^1$, and indeed it turns out that the pair of equations
\begin{eqnarray}
  \eth'\omega^0 &=&0\label{ethomz}\\
  \thorn'\omega^0&=&0 \label{thornomz}
\end{eqnarray}
is BMS-invariant, has a $\C^2$ of solutions (valid over all $\scrif$, not just one generator), and fields satisfying these are the accepted definition of asymptotically constant spinors \cite{PR1986}. 
{\em We now enforce eqs. (\ref{ethomz}), (\ref{thornomz}).}

Each non-zero such field $\omega^0$ will vanish precisely along one generator $\gamma(\omega^0)$ of $\scrif$, and so determines an asymptotic direction.  In fact, one should think of $\omega^0$ as conveying the information of the $\pi_{A'}$ coordinates of the twistor in Minkowski space --- the tangent spinor.  (The mixing of finite space-time $\pi_{A'}$ coordinates and asymptotic $\omega^0$ field is due to the conformal rescaling made in passing to $\scrif$.)  The twistors with $\omega^0$ vanishing identically are a thin set ``at infinity in twistor space,'' and we will not need to consider them; {\em we assume $\omega^0$ does not vanish identically.}

The remaining data for the twistor will be specified on the generator $\gamma (\omega^0)$.  To do this, we must know how to transport $\omega^1$ and ${\tilde\pi}_{0'}$ along the generator.  This is done via {\em twistor transport} (deducible from the twistor equation) \cite{PR1986,ADH2007}, which gives us
\begin{eqnarray}
  \thorn'\omega^1 &=& -i{\tilde\pi}_{1'} \\
       &=&\eth\omega^0\label{trano}\\
  \thorn'{\tilde\pi}_{0'} &=& 0\, .\label{tranone}
\end{eqnarray}  

{\em In sum, then, we specify a twistor by giving a non-zero $\omega^0$ field satisfying eqs. (\ref{ethomz}), (\ref{thornomz}), and then data $\omega^1$, ${\tilde\pi}_{0'}$ satisfying eqs. (\ref{trano}), (\ref{tranone}) along $\gamma(\omega^0)$.}  A twistor is {\em real} if $\omega^1$ vanishes somewhere along $\gamma(\omega^0)$.\footnote{Assuming the generator is infinitely long.  If not, eq. (\ref{trano}) implies the complex-valued field $\omega^1$ is an affine function of the real parameter $u$, and we formally extend this function to all real values of $u$ to see whether it vanishes.}  Such a twistor is identifiable as a null geodesic meeting $\scrif$ at this point; see ref. \cite{ADH2021b} for explicit formulas.  

The result of this is an eight-real-dimensional twistor space $\TT$ which has an invariant existence as a manifold equipped with certain other structures (as we will see, it follows from the above that it is a bundle of affine $\C^2$'s over the space of $\omega^0$ fields).  That structure is weaker than what we would have for special relativity.

On any cut, we may use the twistor data\footnote{In the following expression, I abuse notation slightly by using $\z$ to stand for the cut as a point-set, and not just a function.} $(\omega^A,{\tilde\pi}_{A'})\Bigr|_{\z\cap\gamma(\omega^0)}$ at the point where the cut intersects $\gamma(\omega^0)$ to solve the quasilocal twistor equations, and that solution space is naturally a $\C^4$.  This means that each cut gives us a chart mapping $\TT$ to $\T(\z)\simeq\C^4$, but the patching functions $\C^4\to\C^4$ relating the charts for different cuts are not complex-linear maps.  The twistor space $\TT$ has an invariant existence, but as a manifold, not as a complex vector space.

In special relativity, it is well-understood how to translate twistor structures to space-time ones.  Because in the Bondi--Sachs case the twistor structure is weaker than the Minkowskian one, this process must be revisited in order to understand what the details of the interpretation of the angular momentum, the spin and the center of mass are.  This is done in the references; in what follows I will just explain the ideas behind the computations and give the results.

\subsubsection{Angular momentum}

Penrose's quasi-local formula for the angular momentum is given at $\scrif$ by
\begin{eqnarray}\label{amom}
  A_\z(Z) &=&\frac{1}{8\pi G}\oint_\z \psi_{ABCD}\omega^A\omega^B\epsilon_{C'D'}\, dx^{CC'}\wedge dx^{DD'}\, ,\qquad
\end{eqnarray}
where $\psi_{ABCD}$ is the rescaled Weyl spinor \cite{PR1986}, and $\omega^A$ is the spinor field for $Z$ as a quasilocal twistor at $\z$.
This, then, is interpreted as the angular momentum evaluated at the cut $\z$.  When the twistor is real, it is thought of as before as determining a null geodesic and a tangent spinor, and $A_\z(Z)$ codes the angular momentum with origin that null geodesic and component determined by $\omega^0$.  The twistor $Z$ is not restricted to meet the cut $\z$.

The most important point is that the same twistor space $\TT$ is used no matter what cut is chosen.  Thus angular momenta at different cuts can be directly compared.

\section{Sup-norm continuity of $A_\z(S)$}

The angular momentum at any cut is given by eq. (\ref{amom}).
To understand how this varies with $\z$, we must know the behaviors of $\psi_{ABCD}$, $\omega^A$, and the tangent vectors of $\z$ which will weight the two-form.  Of these, the spinor field $\psi_{ABCD}$ is certainly a sup-norm continuous function of $\z$, and the behavior of the tangents is elementary.  It is the twistor field which requires a substantial computation.

In what follows, I will work in an arbitrary but fixed Bondi frame, keeping standard notation.\footnote{So I will be coordinatizing the twistor space $\TT$ by the chart determined by a cut $u=u_0$ in that frame.}  I will also have to consider quantities associated with the supertranslated cut $u=\z$; I will generally indicate these by postscripts.  For instance, the shear at $\z$ would be $\sigma_\z =\SB(\z ) -\eth^2\z$.  (Note that in the expression just given, the operator $\eth$ is unambiguous, because $\z$ is a function of angle only, but in general we must bear in mind a distinction between $\eth$ and $\eth_\z$.)

\subsection{The tangents and differential on $\z$}

The antiholomorphic tangent for $\z$ will be
\begin{eqnarray}
  m^a_\z &=& m^a +(\eth \z) n^a\, ,
\end{eqnarray}
and the omicron spinor will be
\begin{eqnarray}
  \omicron^A_\z &=& \omicron^A +(\eth\z)\iota^A\, .
\end{eqnarray}
The differential
\begin{eqnarray}    
  (dx^c\wedge dx^d)_\z &=&(2i) m^{[c}_\z{\overline m}^{d]}_\z \, d{\mathcal S}\\
    &=& 2i\left( m^{[c}{\overline m}^{d]} \right.\nonumber\\
    &&\left.
    +(\eth \z) n^{[c}{\overline m}^{d]} 
      +(\eth'\z) m^{[c}n^{d]}\right)\, d{\mathcal S}\, ,\qquad
\end{eqnarray}
where $d{\mathcal S}$ is the standard area measure on the sphere.  Then the quantity entering in the expression (\ref{amom}) for the angular momentum will be
\begin{eqnarray}\label{form}
  \epsilon_{C'D'}  (dx^{CC'}\wedge dx^{DD'})_\z &=&
  -2i (\omicron^{(C}\iota^{D)}    +(\eth\z ) \iota^C\iota^D )\, d{\mathcal S}\, .\qquad
\end{eqnarray}
One notes the potentially problematic derivative $\eth\z$.

\subsection{The twistors}

We recall that the twistor field components $\omega^0$ are well-defined on $\scrif$, but the $\omega^1$ parts are cut-dependent.  

It is known \cite{ADH2007,ADH2021b} that, on the $u=\const$ cuts of our Bondi slicing, we have
\begin{eqnarray}
  \omega^1 &=& \omega^0\eth\lambda -\lambda\eth\omega^0 +\alpha\eth\omega^0
    +\beta\overline{\omega^0}
    \, ,
\end{eqnarray}
where $\lambda$ is any function satisfying $\eth^2\lambda =\SB$, and we have the evolution equations
\begin{eqnarray}
\alpha (u)-\alpha(u_0) &=&u-u_0 +\lambda (u,\gamma ) -\lambda(u_0,\gamma )\\
\beta (u)-\beta(u_0) &=&\left(\frac{\eth\omega^0}{\, \overline{\eth\omega^0}\, } 
  \left( \eth'\lambda (u) -\eth'\lambda(u_0)\right)\right)\Bigr|_\gamma\, ,
\end{eqnarray}
{\em where $\gamma=\gamma (\omega^0)$ is the generator on which $\omega^0$ vanishes.}
The quantity $\lambda$ is an angular potential for the shear, and plays a central role in the treatment of angular momentum.  There is freedom to adjust the $j=0$ and $j=1$ parts of $\lambda$ arbitrarily, but the same effects can be achieved by changing $\alpha$ and $\beta$.  {\em It will be convenient to take $\lambda =\eth^{-2}\SB$, where $\eth^{-2}$ is the Green's operator producing zero $j=0$ and $j=1$ contributions.}

To work out the corresponding field $\omega^1_\z$ on the cut $u=\z$, we first wish to find a corresponding angular potential $\lambda_\z$, so
\begin{eqnarray}\label{skef}
     \eth_\z^2\lambda_\z &=&\sigma_\z\nonumber\\
       &=&\SB (\z )-\eth^2\z\, .
\end{eqnarray}
Here we understand $\SB(\z) =\SB (\z (\gamma ),\gamma)$, where $\gamma$ denotes a point on the sphere of generators of $\scrif$ and $\SB (u,\gamma )$ is the Bondi shear.
The right-hand side of eq. (\ref{skef}) is only a function of angle, so we may take
\begin{eqnarray}\label{shek}
  \lambda_\z &=&\eth^{-2}\SB(\z ) -\z\, .
\end{eqnarray}
The first term on the right of eq. (\ref{shek}) is complicated, owing to the composition of $\SB$ and the cut function $\z$, but the operator $\eth^{-2}$ tends to increase differentiability.  We will see shortly that $\lambda_\z$ is sup-norm continuous in $\z$.

At this point, we know that on $u=\z$ the twistor field
\begin{eqnarray}\label{omeqone}
  \omega^1_\z &=& \omega^0\eth\lambda_\z -\lambda_\z\eth\omega^0 
    +\alpha_\z\eth\omega^0 +\beta_\z\overline{\omega^0}\, ,
\end{eqnarray}
for some constants $\alpha_\z$, $\beta_\z$.  (In principle, all of the $\eth$ operators in eq. (\ref{omeqone}) are really $\eth_\z$, but as they act on $u$-independent quantities there is no ambiguity.)  To work out the constants, we enforce equality of the data $\omega^A_\z=\omega^A$, ${\tilde\pi}^\z_{A'}={\tilde\pi}_{A'}$ at the point $\z\cap\gamma(\omega^0)$.  

We see from eq. (\ref{omonech}) that the equality $\omega^A_\z=\omega^A$ at $\gamma$ is equivalent to requiring $\omega^1_\z =\omega^1$ there.
We have
\begin{eqnarray}
  \omega^1_\z\Bigr|_{\gamma}
    &=& (\alpha_\z -\lambda_\z(\gamma))\eth\omega^0\Bigr|_{\gamma}\, ,
\end{eqnarray}
and the corresponding equation without the $\z$ subscripts.  Equating these and using eq. (\ref{shek})
gives us
\begin{eqnarray}\label{alphaz}
  \alpha_\z 
  &=&\eth^{-2}\SB(\z)\Bigr|_\gamma-u_0 +\alpha (u_0) -\lambda (u_0,\gamma)
  \, .\qquad
\end{eqnarray}      

Similarly, we have from eq. (\ref{pizeroch}), that
\begin{eqnarray}
  {\tilde\pi}_{0'}^\z &=&{\tilde\pi}_{0'} + (\eth'\z){\tilde\pi}_{1'}\\
      &=&{\tilde\pi}_{0'} + i(\eth'\z)\eth\omega^0\label{pieq}
\end{eqnarray}
at $\z\cap\gamma$.   Using
\begin{eqnarray}
 {\tilde\pi}_{0'}^\z &=&
  i\eth'\omega^1_\z\Bigr|_{\gamma}\\
    &=&i\{ -(\eth'\lambda_\z )(\eth\omega^0) +\beta_\z\overline{\eth\omega^0}\}\Bigr|_{\gamma}
\end{eqnarray}
and
\begin{eqnarray}
   {\tilde\pi}_{0'} 
    &=&i\{ -(\eth'\lambda )(\eth\omega^0) +\beta(\z )\overline{\eth\omega^0}\}\Bigr|_{\gamma}
    +i(\eth'\z)\eth\omega^0\Bigr|_{\gamma}
    \, ,\qquad
\end{eqnarray}
in eq. (\ref{pieq}), we find
\begin{eqnarray}\label{betaz}
  \beta_\z 
      &=&\beta(u_0) +\left[\frac{\eth\omega^0}{\, \overline{\eth\omega^0}\, }
      \left(-\eth'\lambda\Bigr|_{ u_0} +\eth'\eth^{-2}\SB(\z) \right)\right]
      \Bigr|_{\gamma}\
    \, .\qquad
\end{eqnarray}

Putting the results (\ref{transom}), (\ref{omzch}), (\ref{shek}), (\ref{omeqone}), (\ref{alphaz}), (\ref{betaz})  into eq. (\ref{omz}), we have the explicit formula
\begin{eqnarray}\label{twexn}
\omega^A_\z &=&\omega^0\omicron^A +{\hat\omega}^1\iota^A\, ,
\end{eqnarray}
where
\begin{widetext}
\begin{eqnarray}\label{twexp}
{\hat\omega}^1=
  \omega^0\eth\eth^{-2}\SB(\z) &+&\left[ (\eth^{-2}\SB(\z)\Bigr|_\gamma)-u_0
         +\alpha(u_0)-\lambda(u_0,\gamma)-\eth^{-2}\SB(\z) +\z\right]\eth\omega^0
         \nonumber\\
         &+&\left[\beta(u_0) +\frac{\eth\omega^0}{\, \overline{\eth\omega^0}\, }\left(
           -\eth'\lambda(u_0) +\eth'\eth^{-2}\SB(\z)\right)\right]\Bigr|_{\gamma}
               \overline{\omega^0}\, .
\end{eqnarray}
\end{widetext}               
We note for later use that
\begin{eqnarray}\label{hepz}
\eth{\hat\omega}^1 &=&\omega^0\SB(\z)
  +(\eth\z)\eth\omega^0\, ,
\end{eqnarray}
in consequence of the relation $\eth^2\omega^0=0$
(holding for any spin-weight $-1/2$ field $\omega^0$ satisfying $\eth'\omega^0=0$).

\subsection{Bundle structure}

Although any fixed cut $\z$ gives twistor space $\TT$ a vector-space structure $\T(\z)$, its invariant structure is weaker.    This means that, while the angular momentum $A_\z$ is a quadratic function relative to the structure $\T(\z )$, it will have a more complicated functional form relative to $\T(\acute\z )$.
So, from an invariant point of view, the function $A_\z$ on $\TT$ must be regarded as lying in some function space.  

While we have more work to do before taking up the function $A_\z$ itself, this is a natural point to clarify the invariant structure of $\TT$, which will be important in specifying the class of functions to be used.

We have been working in a chart derived from the cut $u=u_0$, in which the twistors are coordinatized by $(\alpha (u_0),\beta(u_0))\in\C^2$ and the spinor field $\omega^0$; recall the allowable spinor fields from a two-complex-dimensional vector space, but we exclude zero.  {\em With this understanding, I will write $\omega^0\in\C^2-\{ 0\}$.}  These spinor fields are invariantly defined, and so $\TT$ is invariantly a bundle over them, that is, over $\C^2-\{ 0\}$.

In fact, this is a bundle of affine $\C^2$'s.  This follows from eqs. (\ref{alphaz}), (\ref{betaz}), which show that $(\alpha(u_0),\beta(u_0))$ changes by an $\omega^0$-dependent translation in passing to the chart derived from the cut $u=\z$.

We will see below that $A_\z$ can be regarded as a polynomial in $(\alpha (u_0),\beta (u_0))$, whose coefficients are sections of certain line bundles over the sphere of generators of $\scrif$, the sphere being the space $\C^2-\{ 0\}$ taken projectively.  While the particular polynomials depend on the chart, the class of them (of given degrees) does not.

\subsection{Sup-norm continuity of the twistor fields}

Now let us consider the regularity of the quantities $\eth^{-2}\SB(\z)$, $\eth\eth^{-2}\SB(\z)$, $\eth'\eth^{-2}\SB(\z )$ figuring in the formula (\ref{twexp}) for the twistors.  They will be shown to be jointly continuous in the cut in sup norm and the point at which they are evaluated.  The argument is based on the ellipticity of $\eth$, $\eth'$.  (For analytic background used in the following, see the appendix of Besse \cite{Besse}.)

We will be considering whether $\SB(\z )$ and related quantities lie in $L^p$ spaces on the sphere. 
Now, the Bondi shear $\SB$ is a spin-weight two quantity, and so it takes values in a certain line bundle over the sphere.  There is an inner product on the fibers, so there is a well-defined $L^p$ norm.  (The norm depends on the Bondi frame, but the resulting topology does not.)  We will not need to write this norm explicitly, though, and in the following the notation $|\cdots|$ stands for the ordinary modulus of a complex quantity (or of an element of a line bundle with Hermitian inner product).

We first note that $\SB(\z )$ is a continuous function on the sphere, and hence an element of each $L^p$ space for $p>1$.  The map $\z\mapsto\SB(\z )$ (as an element of the $L^p$ space) is also sup-norm continuous.  To see this, note that since $\SB$ is $C^1$, for $\z$ uniformly close enough to $\z_0$, we will have
\begin{eqnarray}
  |\SB(\z )-\SB(\z_0)| &\leq& M|\z -\z_0|
\end{eqnarray}
pointwise on $S^2$,
where $M$ is a bound on $|\dot\sigma|$ in the corresponding neighborhood of the image of $\z_0:S^2\to\scrif$.  
But then the $L^p$ norm of $\SB(\z )-\SB(\z_0)$ is bounded by $M$ times the sup norm of $\z-\z_0$ in this neighborhood of $\z_0$, and so the map $\z\mapsto\SB(\z)$ is sup-norm continuous to $L^p$.

By elliptic regularity, the quantity $\eth^{-2}\SB(\z)$ will exist as an element of the Sobolev space $W_2^p$, and vary sup-norm continuously with $\z$.  By the embedding theorems, it will exist as an element in a H\"older space (with, in fact, some positive exponent), again varying sup-norm continuously with $\z$.  
Considering now the composition (with ${\mathcal C}$ the space of cuts in sup norm)
\begin{eqnarray}
\begin{matrix}
  S^2\times {\mathcal C} &\to &S^2\times C^0(S^2,\C )&\to &\C\\
  (\gamma ,\z )&\mapsto& (\gamma, \eth^{-2}\SB(\z))&\mapsto&\eth^{-2}\SB(\z)\Bigr|_\gamma \, ,
  \end{matrix}
\end{eqnarray}
where the last step is the (jointly continuous)
evaluation map $S^2\times C^0(S^2,\C)\to \C$, we see $\eth^{-2}\SB(\z)$ is jointly continuous, in the point $\gamma\in S^2$ it is evaluated at and in $\z$ in sup norm.

Since $\lambda_\z=\eth^{-2}\SB(\z)-\z$, this result implies $\lambda_\z$ is jointly continuous.

The only difference in the argument for $\eth\eth^{-2}\SB(\z)$, $\eth'\eth^{-2}\SB(\z)$ is that the extra $\eth$, $\eth'$ derivatives lead to the replacement of the Sobolev space with $W_1^p$.  This is still sufficient for the embedding theorems.

These results imply the twistor fields, eqs. (\ref{twexn}), (\ref{twexp}), are sup-norm continuous.

\subsection{Sup-norm continuity of the angular momentum}

With the regularity of the twistor fields
established, 
we will see
it is a routine matter to substitute the expression (\ref{form}) for the area element and the formula (\ref{twexp}) for the twistors into Penrose's angular momentum (\ref{amom}), and rewrite it by repeated integations by parts to verify its sup-norm continuity.

First, though, we make precise the topology in which the function $A_\z$ on $\TT$ is considered, and then we organize the problem so the essential issue is apparent.

\subsubsection{Nature of the continuity}

The defining formula (\ref{amom}) for the angular momentum, together with the expressions (\ref{twexn}), (\ref{twexp}) for the twistors, show that
$A_\z$ is a polynomial in $\alpha(u_0)$, $\beta(u_0)$, with coefficients which are functions of $\omega^0$.  (Here, as before, we are thinking of the field $\omega^0\in\C^2-\{0\}$.)

The coefficient functions are homogeneous in $\omega^0$ (changing under 
rescalings $\omega^0\mapsto k\omega^0$ by factors $k^n{\overline k}^{2-n}$ for $n=0,1,2$), so they can be regarded as functions on the sphere taking values in certain line bundles.  
As is standard, to deal with the homogeneity, we will divide by a Hermitian norm $\|\omega^0\|^2$ on the space of allowable spinor fields.  Then the normalized modulus $|A_\z(Z)|/\|\omega^0\|^2$ is independent of the scale of $\omega^0$, and each coefficient function, divided by $\|\omega^0\|^2$, will have a well-defined sup norm over the sphere, which we use the topologize the function space.   (The topology is independent of the Hermitian norm $\|\omega^0\|$.)  So the angular momentum map $\z\mapsto A_\z$, from cuts to our functions on the sphere, will be analyzed with sup-norm topology on both source and target.

As noted above, although the particular coefficient functions will change as the chart changes, the class of line bundles which arises is invariant.

\subsubsection{Criterion for reducing the problem}

The essential step will be to show that each of these normalized coefficient functions can be reduced to a sum of integrals of the form
\begin{eqnarray}\label{cumb}
\oint f(\z , \eth^{-2}\SB(\z), \eth\eth^{-2}\SB(\z),\eth'\eth^{-2}\SB(\z),\omega^0,\eth\omega^0,\gamma )  \, d{\mathcal S}\, ,\qquad
\end{eqnarray} 
where no derivative of $\z$ appears in the list of arguments.
The function $f$ is continuous and changes under rescalings of $\omega^0$ by at most a phase.  

Writing now briefly $I(\z,\omega^0)$ for the expression (\ref{cumb}) (the integral, {\em not} the integrand), the quantity
\begin{eqnarray}\label{cuke}
  \left| I(\z,\omega^0) -I(\acute\z,\omega^0)
    \right|
\end{eqnarray}
will be a function on the sphere, whose supremum
can be made as small as desired by requiring $\acute\z$ be close to $\z$ in sup norm, because the function $f$ in eq. (\ref{cumb}) is continuous and the modulus (\ref{cuke}) depends only projectively on $\omega^0$ (that is, depends only on $\gamma\in S^2$).   This will establish the required continuity.  So the key thing will be to rewrite the integrands so they do not depend on derivatives of $\z$.

\subsubsection{Final step}  

We will now show that by repeated integrations by parts we can write the twistor angular momentum $A_\z(Z)$ as a sum of explicitly sup-norm continuous terms.
These parts-integrations are over the cut $\z$, and are in terms of the eth operator $\eth_\z$ tangent to that cut.  
This is related to the operator $\eth$ of the Bondi system by $\eth_\z =\eth +(\eth\z)\thorn'$, where $\thorn'$ is the derivative in the $u$-direction.  We must take this into account.

The quantities which will enter are the curvature spinor $\psi_{ABCD}$, the spinor dyad, and the twistor fields $\omega_\z^A$.  Of these, there is no real issue of distinction as far as the operators' actions on the dyad and the twistor fields go, since the dyad spinors are parallel-transported along the generators of $\scrif$ and the twistor fields were defined as functions of angle only.  
The curvature spinor, however, is certainly not parallel-transported along the generators, and we must distinguish between $\eth$ and $\eth_\z$ acting on it.

Write the angular momentum as a sum of two terms:
\begin{eqnarray}
A_\z(Z) &=& A_1+A_2\, ,
\end{eqnarray}
with\footnote{For the reader examining the details of these formulas in terms of the expressions (\ref{twexn}), (\ref{twexp}), for the twistor fields, it is perhaps helpful to emphasize that $A_\z(Z)$ is a function of the choice of field $\omega^0 \in\C^2-\{ 0\}$, $\alpha(u_0)$, $\beta(u_0)$.  Inside the integrals, the form of $\omega^0$ as a function of the point of integration on the sphere enters.  In the formula (\ref{twexp}) for ${\hat\omega}^1$, some quantities are evaluated at $\gamma$; recall that this is the generator at which $\omega^0$ vanishes, so $\gamma$ does not vary with the point of integration.} 
\begin{eqnarray}
  A_1&=&\frac{-i}{4\pi G}\oint_\z \psi_{ABCD}\omega^A_\z \omega^B_\z
    \omicron^C\iota^D\, d{\mathcal S}\label{A1eq}\\
  A_2&=&\frac{-i}{4\pi G}\oint_\z \psi_{ABCD}\omega^A_\z \omega^B_\z
    \iota^C\iota^D\eth\z\, d{\mathcal S}\, .\label{A2eq}
\end{eqnarray}
The term $A_1$
is evidently sup-norm continuous in $\z$.  
The term $A_2$
is not obviously so.  However we may regard the integrand in eq. (\ref{A2eq}) as
\begin{eqnarray}
  (\eth_\z -\eth)\psi^{(-1)}_{ABCD}\omega_\z^A\omega_\z^B\iota^C\iota^D\, ,
\end{eqnarray}
where  
$\psi^{(-n)}_{ABCD}$ is an $n^{\rm th}$ $u$-antiderivative of $\psi_{ABCD}$.  
On integrating, the total $\eth_\z$ derivative will vanish, and we are left with
\begin{eqnarray}
 A_2 &=&A_{21}+A_{22}\, ,
\end{eqnarray}
where
\begin{eqnarray}
  A_{21}&=&\frac{+i}{4\pi G}\oint_\z (\eth\psi^{(-1)}_{ABCD})
       \omega_\z^A\omega_\z^B\iota^C\iota^D\, d{\mathcal S}\\
 A_{22}&=&\frac{+2i}{4\pi G}\oint_\z \psi^{(-1)}_{ABCD}
       \omega_\z^A(\eth\omega^0)(\omicron^B+(\eth\z)\iota^B)\iota^C\iota^D\, d{\mathcal S}\, .\qquad
\end{eqnarray}
Here $A_{21}$ is manifestly sup-norm continuous.

We now iterate this procedure.  We have
\begin{eqnarray}
  A_{22}&=&A_{221}+A_{222}\, ,
\end{eqnarray}
where
\begin{eqnarray}
 A_{221}&=&\frac{+2i}{4\pi G}\oint_\z \psi^{(-1)}_{ABCD}
       \omega_\z^A(\eth\omega^0)\omicron^B\iota^C\iota^D\, d{\mathcal S}
         \\
A_{222}&=&\frac{+2i}{4\pi G}\oint_\z \psi^{(-1)}_{ABCD}
       \omega_\z^A(\eth\omega^0)(\eth\z)\iota^B\iota^C\iota^D\, d{\mathcal S}\, .
\end{eqnarray}
Here $A_{221}$ is manifestly sup-norm continuous, and we may rewrite
\begin{eqnarray}
  A_{222} &=&\frac{+2i}{4\pi G}\oint_\z
    (\eth_\z -\eth) (\psi^{(-2)}_{ABCD}\omega^A_\z (\eth\omega^0)\iota^B\iota^C\iota^D)\, d{\mathcal S}
      \nonumber\\
    &=& A_{2221}+A_{2222}\, ,
\end{eqnarray}
with
\begin{eqnarray}
  A_{2221}  &=&\frac{-2i}{4\pi G}\oint_\z
     (\eth\psi^{(-2)}_{ABCD})\omega^A_\z (\eth\omega^0)\iota^B\iota^C\iota^D\, d{\mathcal S}\\  
  A_{2222} &=&\frac{-2i}{4\pi G}\oint_\z
   \psi^{(-2)}_{ABCD}(\omicron^A+(\eth\z)\iota^A) (\eth\omega^0)^2\iota^B\iota^C\iota^D\, d{\mathcal S}\, .
   \qquad
\end{eqnarray}
Again, the first term $A_{2221}$ is sup-norm continuous, and the second can be written as a sum
\begin{eqnarray}
  A_{2222}&=& A_{22221}+A_{22222}\, ,
\end{eqnarray}
where
\begin{eqnarray}  
  A_{22221} &=&\frac{-2i}{4\pi G}\oint_\z
   \psi^{(-2)}_{ABCD}\omicron^A (\eth\omega^0)^2\iota^B\iota^C\iota^D\, d{\mathcal S}\\
  A_{22222}&=&\frac{-2i}{4\pi G}\oint_\z
   \psi^{(-2)}_{ABCD}(\eth\z)\iota^A(\eth\omega^0)^2\iota^B\iota^C\iota^D\, d{\mathcal S}\, .\qquad
\end{eqnarray}
Again, the first is sup-norm continuous, and we can apply the trick again to the second, getting
\begin{eqnarray}
  A_{22222} &=& \frac{-2i}{4\pi G}\oint_\z (\eth_\z -\eth)
     (  \psi^{(-3)}_{ABCD}\iota^A(\eth\omega^0)^2\iota^B\iota^C\iota^D)\, d{\mathcal S}\nonumber \\
     &=&\frac{+2i}{4\pi G}\oint_\z(\eth\psi^{(-3)}_{ABCD})\iota^A(\eth\omega^0)^2\iota^B\iota^C\iota^D
       \, d{\mathcal S}\, ,
\end{eqnarray}
which is manifestly sup-norm continuous.
 
This concludes the proof that the twistor angular momentum $A_\z(Z)$ is sup-norm continuous.

\section{Spin and center of mass}

As noted in the introduction, when we use a natural general-relativistic extension of the twistor formulas for spin and center of mass, we find that these are not simply the two $j=1$ parts of some spinor or tensor field, but contain higher-$j$ contributions, which have important geometric interpretations.  In the case of the center of mass, they account for possible deformations of the active cut $\z$ from a physically sensible center of mass; for spin, an analogous result holds, with the Newman--Winicour interpretation of spin as a displacement of the center of mass into the complex.

To write these formulas in an accessible fashion, I will represent asymptotically constant spinors by boldface symbols like $\bm{\pi_{A'}}$ (rather than fields like $\omega^0$ --- see ref. \cite{ADH2021b} for an explicit dictionary, and for derivations of the following formulas).  The quantity
\begin{eqnarray}
  2i\bm{\mu^{A'B'}\pi_{A'}\pi_{B'}} =A_\z(Z)\, ,
\end{eqnarray}
where $Z$ is a twistor with $\alpha_\z=\beta_\z=0$ and $\omega^0$ corresponding to $\bm{\pi_{A'}}$.  Then the spin is
\begin{eqnarray}\label{spin}
\bm{J} (\gamma )
 &=& \Im\left( -2\bm{\mu^{A'C'}t^A{}_{C'}}-M\lambda_\z(\gamma)\bm{t^{AA'}}\right)
   \bm{{\overline\pi}_A\pi_{A'}}\, ,\qquad
\end{eqnarray}
where $\bm{t^a}$ is a unit vector along the direction of the Bondi--Sachs energy-momentum, and $M$ is the mass.  Here we think of the null vector $\bm{{\overline\pi}_A\pi_{A'}}\leftrightarrow \gamma$ as determining a direction on the sphere, so $\bm{J}$ is a function of this direction.
The first term on the right of eq. (\ref{spin}) (after distributing over the subtraction)
is the $j=1$ contribution familiar from special relativity, and the second, involving $\Im\lambda_\z$, has the $j\geq 2$ general-relativistic corrections.  

The center of mass is given by the cut
\begin{eqnarray}
  \z_{\rm cm} &=& \z +\Re\lambda_\z
   +2\frac{\Re\bm{\mu^{A'B'}t^A{}_{C'}{\overline\pi}_A\pi_{A'}}}{M\bm{t^{AA'}{\overline\pi}_A\pi_{A'}}}
   +\const\, .\qquad
\end{eqnarray}
Here the last term (before ``$+\,\const$'') has the same form as for Minkowski space, and is the $j=1$ contribution, and the $\Re\lambda_\z$ is a correction term which removes any possible gauge effects.
The freedom to add a constant is expected:  in special relativity, the center of mass is a world-line.  If desired, the constant here could be adjusted so that $\z_{\rm cm}-\z$ is minimal in $L^2$ (with respect to the frame determined by $\bm{t^a}$); this would in a natural sense be the center of mass closest to $\z$.

Our results above show these expressions are sup-norm continuous:  formulas (\ref{alphaz}), (\ref{betaz}) show that the conditions $\alpha_\z=0$, $\beta_\z=0$ lead to sup-norm continuous values of $\alpha(u_0)$, $\beta(u_0)$, and so
$2i\bm{\mu^{A'B'}}$ is sup-norm continuous, and we have seen $\lambda_\z$ is sup-norm continuous.
(Since the Bondi--Sachs energy-momentum admits a flux density, we have $M$ and $\bm{t^a}$ sup-norm continuous.)

\section{Quasi\-conventional representation}

Twistor theory defines angular momentum as a function on twistor space, not a field on space-time (or on some model space-time), and it also leads to geometric definitions of the center of mass and spin.  
We have seen that all of these things are sup-norm continuous.

It is possible --- and, in my view, the successes of the twistor program make it likely --- that we will come to regard this twistor perspective as the best way of thinking of angular momentum for Bondi--Sachs space-times.  On the other hand, it {\em is} unconventional, and so one does want to connect it more explicitly to conventional treatments.  I explain here how this may be done, and what continuity properties it has.

In Minkowski space, to recover from twistors the angular momentum as a spinor or tensor field, one simply restricts the twistor $Z$ in $A(Z)$ to be a null geodesic through the space-time point of interest.  In the Bondi--Sachs context, however, there is no compelling structure at $\scrif$ replacing the the Minkowski space-time points.  We may however do something like what is done in more conventional approaches, and select a cut $\z_{\rm pas}$, and then choose the twistors to range over the null geodesics meeting $\z_{\rm pas}$ orthogonally.  This would reduce to the standard result for $\z_{\rm pas}$ a good cut in a Minkoskian regime (and $\z_{\rm act}$ any cut in that regime).   
I will call the resulting representation of the angular momentum {\em quasi\-conventional.}

While the quasi\-conventional representation is evidently in a general sense related to conventional approaches, there are significant differences both in the quantities and in their interpretational frameworks.

A few points are worth making:

(a) The condition that $Z$ meet $\z_{\rm pas}$ orthogonally will bring in the $C^1$ structure of $\z_{\rm pas}$, but not $\z_{\rm act}$.
The cut-continuity criterion proposed by Chen et al. would apply to deformations of the {\em active} cut $\z_{\rm act}$, holding the passive cut fixed.  This is because the question to be addressed is how the angular momentum varies with the choice of active cut.

Chen et al. consider angular momenta indexed by elements of the BMS algebra, and do so ``weakly,'' that is, holding the BMS algebra element fixed but allowing the cut to vary.  The passive cut in this case is the origin of the Poincar\'e subgroup the BMS element generates, and this cut must be at least $C^1$ if the BMS vector field is to be $C^0$.  Thus the BMS-based approaches, at least in their most straightforward interpretations, require at least as much regularity as the twistor one.  (In fact, one would most naturally require the BMS generators to preserve whatever regularity structure is assumed for $\scrif$, and this would mean more regular passive cuts.)

(b) We expect the quasi\-conventional angular momentum to be interpretable as a mild deformation of a special-relativistic construction only to the degree which $\z_{\rm pas}$ is in some suitable sense close to a good cut.  For other choices, the construction is more formal and its interpretation is less clear.  Although this is a negative statement, it can be regarded as clarifying the function of the passive cut and the transition from weakly to a strongly general-relativistic circumstances.  (It is worth thinking about parallel implications for conventional treatments; I will discuss this below.)

(c) As the twistor $Z$ varies over the null geodesics meeting $\z_{\rm pas}$ orthogonally, we get a (spin-weight $-1$) function on the sphere of generators.  In the special-relativistic limit, with $\z_{\rm pas}$ a good cut, this would simply be the complex $j=1$ representation of the angular momentum, that is, the representation of $\mu^{A'B'}$ as the function $2i\mu^{A'B'}\pi_{A'}\pi_{B'}$ for $\pi_{A'}$ in the spin-bundle over the sphere.  In the general-relativistic case, however, there will usually be $j\geq 2$ contributions as well.  These are related to the contributions of $\lambda$ to the center of mass and spin.  (But to extract the center of mass and spin, the twistor approach [Section IV] is more direct.)

I now verify that the quasi\-conventional representation of the angular momentum is, for any fixed $C^1$ passive cut $\z_{\rm pas}$, sup-norm continuous in the active cut $\z_{\rm act}$.  This quantity is
\begin{eqnarray}
  A_{\z_{\rm act}}(Z)\, ,
\end{eqnarray}
where $Z$ varies over the twistors meeting $\z_{\rm pas}$ orthogonally.  Since we already know $A_{\z_{\rm act}}$ is sup-norm continuous in $\z_{\rm act}$, we must show that the restriction of the twistors $Z$ introduces no difficulties.  

However, this restriction is entirely independent of $\z_{\rm act}$.
The conditions for a twistor $Z$ to meet a cut $\z_{\rm pas}$ orthogonally are \cite{ADH2021b}
\begin{eqnarray}
  \alpha (\z_{\rm pas}(\gamma )) &=&\lambda (\z_{\rm pas}(\gamma),\gamma)\\
  \beta (\z_{\rm pas}(\gamma )) &=&-\left( \frac{\eth\omega^0}{\,\overline{\eth\omega^0}\,} 
   \eth' (\z_{\rm pas}-\lambda )\right)\Bigr|_\gamma\, .
\end{eqnarray}   
These imply
\begin{eqnarray}
  \alpha(u_0) 
     &=&-\z_{\rm pas}(\gamma) +u_0 
     +\lambda (u_0,\gamma) \\
  \beta (u_0) 
  &=&-\left(\frac{\eth\omega^0}{\,\overline{\eth\omega^0}\,} \left[\eth' \z_{\rm pas}
   -\eth'\lambda (u_0,\gamma)\right]
  \right)\Bigr|_\gamma\, .
\end{eqnarray}  
We see explicitly that these conditions do not bring in $\z_{\rm act}$ at all.

\section{Flux and flux density}

There is no standard terminology for the various flux quantities of interest.
The mathematically strongest concept I will call {\em flux density,}  
a preferred three-form on $\scrif$, whose integral between two cuts would be the emitted angular momentum.  
Next,
a {\em flux} is the angular momentum emitted between two infinitesimally separated cuts.  Still weaker would be a {\em flux with resect to Bondi parameter,} the angular momentum emitted between two infinitesimally separated $u=\const$ cuts of the same Bondi system.

The flux of the twistor angular momentum with respect to Bondi parameter was derived in ref. \cite{ADH2007}.  We can derive an expression for the flux itself by functionally differentiating the formulas of Section IIIE for $A_\z(Z)$ with respect to $\z$, and also do a related calculation to show that no flux density can exist.

\subsection{Flux}

In the generality of Section IIIE, which considers a cut $\z$ arbitrarily supertranslated with respect to the Bondi coordinates, the expression for the flux is very long.  However, we should not usually need this generality.  It is more natural for most purposes to
ask for the flux between a $u=u_0=\const$ cut of the Bondi system and an infinitesimally separated cut $u=u_0+\delta\z$.  Then we have considerable simplifications.
Using formulas (\ref{A1eq}), (\ref{A2eq}), the flux of angular momentum emitted over $\delta\z$ is (minus)
\begin{eqnarray}\label{fluxone}
\delta A_\z(Z) 
  &=&
\frac{-i}{4\pi G}\oint \{ [
  (\delta\z \thorn'\psi_{ABCD})\omega^A_{u_0}\omega^B_{u_0}\nonumber\\
  &&
    +2\psi_{ABCD}\omega^A_{u_0} \delta\omega^1 \iota^B\} \omicron^C\iota^D]\nonumber\\
    &&-\delta\z\eth(\psi_{ABCD}\omega^A_{u_0}\omega^B_{u_0})\iota^C\iota^D\}\, 
    d{\mathcal S}
    \, ,
\end{eqnarray}
where
\begin{widetext}
\begin{eqnarray}\label{fluxtwo}
\delta\omega^1 &=&
  \omega^0\eth\eth^{-2}({\dot\sigma}_{\rm B}\delta\z )
  +\left[ \eth^{-2}({\dot\sigma}_{\rm B}\delta\z )\Bigr|_\gamma -\eth^{-2}({\dot\sigma}_{\rm B}\delta\z )
    +\delta\z\right]\eth\omega^0
    +\left[ \frac{\eth\omega^0}{\,\overline{\eth\omega^0}\,}\eth'\eth^{-2}({\dot\sigma}_{\rm B}\delta\z )
      \right]\Bigr|_\gamma \overline{\omega^0}\, .
\end{eqnarray}
\end{widetext}      

\subsection{Flux density}

One can also show that a flux {\em density} cannot exist.  The computation is long and the specifics of the results, while having interesting structure, are too long and technical to be worth discussing in detail here.
I will just explain the idea, and it will be seen that the result is expected from the mathematical structure.  It will turn out to have an important 
physical interpretation.  

For a flux density to exist, for any open subset $R_1$ of a cut $\z$ and any first-order perturbation $\delta_1\z$ of $\z$ supported on $R_1$, one should be able to assign a flux $\Phi (\delta_1\z)$ which is linear in $\delta_1\z$, and unaffected by changes in $\z$ whose supports are disjoint from $R_1$.  In particular, then, if we contemplate another perturbation $\delta_2\z$ supported within an open set $R_2$ disjoint from $R_1$, we should have 
\begin{eqnarray}\label{sfc}
  \frac{\delta^2}{\delta_1\z\delta_2\z} A_\z (Z) =0\, ,
\end{eqnarray}
where the left-hand side is the functional derivative.  Equivalently, the kernel
\begin{eqnarray}\label{sfd}
  \frac{\delta^2}{\delta\z (\gamma_1)\delta\z (\gamma_2)} A_\z(Z)
\end{eqnarray}
should have support (in the distributional sense) only on the diagonal $\gamma_1=\gamma_2$.  
(This argument, connecting flux densities to diagonal support of second variations, would apply to any kinematic quantity defined on cuts of $\scrif$.)

However, the quantity (\ref{sfd}) can be computed explicitly (it is easiest again to work around a $u=\const$ cut), and, if 
gravitational radiation ${\dot\sigma}_{\rm B}\not=0$ is present at $\z$, generically has off-diagonal contributions.    These come, for instance, from terms which are products of $\eth^{-2}({\dot\sigma}_{\rm B}\delta_1\z)\Bigr|_\gamma$ with integrals involving curvature terms and $\delta_2\z$, and vice versa.  The first factor involves the Green's function $\eth^{-2}$, so such a term is {\em a product of integrals} over the cut, not a single integral of a product:  they are highly non-diagonal.  In a non-radiating regime, the angular momentum is strictly independent of the cut; contrapositively, the terms here which {\em are} sensitive to the off-diagonal contributions also involve quantities related to the news ${\dot\sigma}_{\rm B}$. 

The non-diagonal contributions to the change in angular momentum (\ref{sfd}) have a direct and important physical interpretation.  They mean that the total angular momentum depends on {\em correlations} of gravitational radiation from separated asymptotic directions (separated generators of $\scrif$).  So the twistor definition conveys very different information, in the radiative case, from one admitting a flux density (such as that of Dray and Streubel).

\section{Discussion}

Chen et al. proposed that any definition of angular momentum for Bondi--Sachs space-times should be sup-norm continuous.  I have shown that this property holds for the twistorial angular momentum, spin, center of mass, and quasi\-conventional representation.  The arguments
turned on the ellipticity of the Newman--Penrose operator $\eth$.  This is also at the heart of the Bondi--Sachs schema, for this operator codes the complex structure of the sphere of generators which is used essentially in defining the Bondi frame, the shear and the news. 

Along the way, and consequently, a number of related results were found.
The formulas (\ref{twexn}), (\ref{twexp}) for the twistors at cuts arbitrarily supertranslated with respect to the working Bondi system
are basic structural results for the theory.  Another is the flux law, eqs. (\ref{fluxone}), (\ref{fluxtwo}), giving the angular momentum emitted between two infinitesimally (but perhaps also supertranslationally) separated cuts.  

Beyond these, there are two points worth discussing in a bit more detail.
One is the lack of existence of a flux density, and the cause of that in the twistor angular momentum's dependence on correlations between the gravitational radiation in different asymptotic directions.
The other is the relation between the twistor definition and BMS-algebra based ones.

\subsection{Flux density versus correlations}

I distinguished in Section VI between the existence of a {\em flux,} the angular momentum emitted between two infinitesimally separated cuts, and a {\em flux density,} a three-form whose integral over the infinitesimal strip separating such cuts would give a flux.  For the twistor angular momentum, a flux exists, essentially in consequence of the sup-norm continuity of the definition, but a flux density does not.

The argument for this uncovered an important point.  A flux density cannot exist if the second variation (\ref{sfd}) of the angular momentum with respect to the active cut has any support off the diagonal.  This off-diagonal support in turn means that there are contributions to the emitted angular momentum which arise from correlations of radiative data in different asymptotic directions.  It is ultimately due to the occurence of the potential $\lambda=\eth^{-2}\SB$, which $\SB$ influences nonlocally, in the twistor formulas.  Most suggested definitions, cast as integrals over the cut involving $\SB$ only locally, will not have this property.\footnote{The CWWY proposal also involves (the real part of) $\lambda$, and in fact the CWWY ``correction term'' with $\lambda$ is one of the terms occurring in the twistor expression.} 

This shows that the twistor angular momentum measures physical effects qualitatively beyond those which can be detected by approaches (such as that of Dray and Streubel) which do admit flux densities. 
It suggests a way of comparing the physical utility of the approaches, by looking at systems (possibly composed of subsystems) which emit radiation in separated directions.  One would look for cases in which one could on clear physical grounds identify the angular momentum emitted, and see whether such correlations were needed to explain it.

It is worthwhile comparing this with the situation for the Bondi--Sachs energy-momentum.  

While the energy-momentum flux density does indeed represent a sort of local structure, there is an important sense in which this locality is relative:  the flux density depends on the news, and the news is not a locally determined quantity --- one gets it from the curvature by doing a non-local angular integral (or a retarded-time integral relative to some initial data).  In other words, once the whole structure of $\scrif$ is known, then the flux density has a well-defined existence:  but if only a portion of $\scrif$ limited in angle (and retarded time) is known, one cannot get the flux density.

The Bondi--Sachs energy-momentum
is an element in the space of asymptotically constant covectors, and to define it we need first to understand the asymptotically constant vectors.  They are constructed, like the twistors, by the use of elliptic systems of equations, which rely on nonlocal structure for their solution \cite{ADH2014}.  Unlike the twistors, though, the asymptotically constant vector fields, once obtained, are well-defined over all of $\scrif$.  It is this which is responsible for the existence of a flux density for the energy-momentum \cite{PR1986}.

Finally, it should be emphasized that these have been questions of nonlocality {\em at scri.}  Nonlocality in this sense is far more severe than in the ordinary one.  For instance, the nonlocalities for binding energies are set by the separation between the systems contributing, but for the geometry of $\scrif$ one it is the distance to the radiation zone.

\enlargethispage{1.5\baselineskip}

\subsection{The interpretation of passive cuts}

To make a link between the twistor approach and others, I introduced the quasi\-conventional representation, which gave the angular momentum 
at any ($C^1$) passive cut.  Yet this resulted in properties different from other approaches:  
the twistorial angular momentum had $j\geq 2$ contributions.  What underlies these differences between the twistor and the BMS-algebra-based approaches?

It comes down to the question of what is taken to code the sense of ``origins'' associated with angular momentum.  In approaches like the Dray--Streubel one, it is simply the passive cuts.  
For twistors, though, the origin information is ultimately in the twistors, and so it is not just the location of a passive cut which matters, but the null geodesics extending orthogonally inwards from it.

In the twistor approach, it is clear that the passive cut is not a compelling or fundamental structure.  It is just a convenience, to try to express the results in a form which can be viewed as a deformation of the special-relativistic structure, and that deformation becomes more severe with the badness of the cut.  If, in some regime, cuts which are not too bad exist, one can think of the quasiconventional $j=1$ part of the angular momentum as a familiar extension of the special-relativistic concept, and the $j\geq 2$ parts as 
extra, general-relativistic portions, with respect to these cuts.  But if all cuts in the regime are pretty bad, then the twistorial interpretation would be that the system is sufficiently general-relativistic that its energy-momentum and angular momentum cannot be well modeled there by something close to a Poincar\'e structure.  The quasi\-conventional representation then has a largely formal role, and to extract information of clear significance one 
cannot avoid the twistorial definition itself.

By contrast, in the BMS-based approaches, the passive cuts have, on their own, essentially no geometric significance.  What is important about them are their relations to the active cuts under consideration --- which active cuts they are supertranslated relative to, and by how much.  

Although the BMS-based {\em formalism} works perfectly mathematically for arbitrary active and passive cuts, the difficulties with center of mass show that even in a Minkowskian regime a bad active cut  is hard to {\em interpret} without bringing in ``by hand'' the regime's good cuts.  In generic circumstances, we have, so far, essentially no guidance in going beyond the formalism to interpretation.

\begin{acknowledgments}

I thank Robert Wald for drawing the paper \cite{CPWWWY} to my attention, and Carlo Morpurgo for useful discussion.  Of course, any errors are my own.

\end{acknowledgments}


%

\end{document}